\documentclass{aa}
\usepackage{lscape}
\usepackage{graphicx}
\usepackage{txfonts}
\usepackage{longtable}
\usepackage{natbib}
\usepackage{url}
\usepackage{float}
\usepackage{aalongtable}
\usepackage{textcomp}
\usepackage{hyperref}
\begin{document}

\title{The CARMENES search for exoplanets around M dwarfs}  
\subtitle{Photospheric parameters of target stars from high-resolution spectroscopy. II. \\
Simultaneous multiwavelength range modeling of activity insensitive lines}
\titlerunning{Simultaneous multiwavelength range modeling of activity insensitive lines}
\author{V.\,M.~Passegger\inst{1}
        \and
        A.~Schweitzer\inst{1}
        \and
        D.~Shulyak\inst{2}
        \and
        E.~Nagel\inst{1}
        \and
        P.\,H.~Hauschildt\inst{1}
        \and
        A.~Reiners\inst{3}
        \and
        P.\,J.~Amado\inst{4}
        \and
        J.\,A.~Caballero\inst{5}
        \and 
        M.~Cort\'es-Contreras\inst{5}
        \and 
        A.\,J.~Dom\'inguez-Fern\'andez\inst{6}
        \and
        A.~Quirrenbach\inst{7}
        \and
        I.~Ribas\inst{8,9}
        \and 
        M.~Azzaro\inst{10}
        \and
        G.~Anglada-Escud\'e\inst{4,11}
        \and
        F.\,F.~Bauer\inst{4}
        \and
        V.\,J.\,S.~B\'ejar\inst{12,13}
        \and
        S.~Dreizler\inst{3}
        \and 
        E.\,W.~Guenther\inst{14}
        \and
        T.~Henning\inst{15}
        \and
        S.\,V.~Jeffers\inst{3}
        \and 
        A.~Kaminski\inst{7}
        \and 
        M.~K\"urster\inst{15}
        \and 
        M.~Lafarga\inst{8,9}
        \and
        E.\,L.~Mart\'in\inst{5}
        \and
        D.~Montes\inst{6}
        \and
        J.\,C.~Morales\inst{8,9}
        \and 
        J.\,H.\,M.\,M.~Schmitt\inst{1}
        \and
        M.~Zechmeister\inst{3}
        }

\institute{
        Hamburger Sternwarte, Gojenbergsweg 112, D-21029 Hamburg, Germany \newline
        \email{vpassegger@hs.uni-hamburg.de}
          \and
          Max Planck Institute for Solar System Research, Justus-von-Liebig-Weg 3, D-37077 G\"ottingen, Germany
          \and
          Institut f\"ur Astrophysik, Georg-August-Universit\"at, Friedrich-Hund-Platz 1, D-37077 G\"ottingen, Germany 
          \and
          Instituto de Astrof\'isica de Andaluc\'ia (IAA-CSIC), Glorieta de la Astronom\'ia s/n, E-18008 Granada, Spain
          \and
          Centro de Astrobiolog\'ia (CSIC-INTA), ESAC, Camino Bajo del Castillo s/n, E-28692 Villanueva de la Ca\~nada, Madrid, Spain
          \and
          Departamento de F\'{\i}sica de la Tierra y Astrof\'{\i}sica and IPARCOS-UCM (Instituto de F\'{\i}sica de Part\'{\i}culas y del Cosmos de la UCM),
          Facultad de Ciencias F\'{\i}sicas, Universidad Complutense de Madrid, E-28040 Madrid, Spain 
          \and
          Landessternwarte, Zentrum f\"ur Astronomie der Universt\"at Heidelberg, K\"onigstuhl 12, D-69117 Heidelberg, Germany
          \and
          Institut de Ci\`encies de l'Espai (CSIC-IEEC), Campus UAB, c/ de Can Magrans s/n, E-08193 Bellaterra, Barcelona, Spain 
          \and
          Institut d'Estudis Espacials de Catalunya (IEEC), E-08034 Barcelona, Spain
          \and
          Centro Astron\'omico Hispano-Alem\'an (CSIC-MPG), Observatorio Astron\'omico  de  Calar  Alto,  Sierra  de  los  Filabres, E-04550  G\'ergal, Almer\'ia, Spain
          \and
          School of Physics and Astronomy, Queen Mary, University of London, 327 Mile End Road, London, E1 4NS, United Kingdom
          \and
          Instituto de Astrof\'{\i}sica de Canarias, c/ V\'ia L\'actea s/n, E-38205 La Laguna, Tenerife, Spain 
          \and
          Departamento de Astrof\'{\i}sica, Universidad de La Laguna, E-38206 La Laguna, Tenerife, Spain 
          \and
          Th\"uringer Landessternwarte Tautenburg, Sternwarte 5, D-07778 Tautenburg, Germany
          \and
          Max-Planck-Institut f\"ur Astronomie, K\"onigstuhl 17, D-69117 Heidelberg, Germany 
          }

\date{}

\abstract
{We present precise photospheric parameters of 282 M dwarfs determined from fitting the most recent version of PHOENIX models to high-resolution CARMENES spectra in the visible (0.52--0.96\,$\mu$m) 
and near-infrared wavelength range (0.96--1.71\,$\mu$m). 
With its aim to search 
for habitable planets around M dwarfs, several planets of different masses have been detected. The characterization of the target sample is important for the ability to derive and 
constrain the physical 
properties of any planetary systems that are detected. As a continuation of previous work in this context, we derived the fundamental stellar parameters effective temperature, surface gravity, and 
metallicity of the CARMENES M-dwarf targets from PHOENIX model fits using a $\chi^2$ method. We calculated updated PHOENIX stellar atmosphere models that include a new equation of state to especially 
account for spectral features of low-temperature stellar atmospheres as well as new atomic and molecular line lists. We show the importance of selecting magnetically insensitive lines for fitting to 
avoid effects of stellar activity in the line profiles. For the first time, we directly compare stellar parameters derived from multiwavelength range spectra, simultaneously observed for the same star. 
In comparison with literature values we show that fundamental parameters derived from visible spectra and visible and near-infrared spectra combined are in better agreement than those derived from the same 
spectra in the near-infrared alone.}
%
 \maketitle
%

\section{Introduction}

In the last decade M dwarfs enjoyed increasing popularity regarding exoplanet surveys. Due to their smaller masses and radii, compared to Sun-like stars, it is easier to detect orbiting planets with the 
transit and radial velocity methods. 
The CARMENES (Calar Alto high-Resolution search for M dwarfs with Exo-earths with Near-infrared and optical \'Echelle Spectrographs) instrument was built to search for Earth-like planets 
in the habitable zones of M dwarfs using the radial velocity technique. CARMENES is mounted on the Zeiss 3.5\,m telescope at Calar Alto Observatory, located in Almer\'{\i}a, in southern Spain 
and has been taking data since January 2016. 
The instrument comprises two fiber-fed spectrographs covering the visible (VIS) and near-infrared (NIR) wavelength regime, from 520 to 960\,nm and from 960 to 1710\,nm with 
spectral resolutions of $R \approx$ 94\,600 and 80\,500, respectively. With simultaneous observations in two wavelength ranges it is easier to identify false positive planetary signals caused by stellar 
activity. Each spectrograph is designed to perform high-accuracy radial velocity measurements with a long-term stability \citep[][]{Quirrenbach2018,Reiners2018a}. 
Several exoplanets have already been detected with CARMENES. The Neptune-mass planets HD~147379~b 
\citep{Reiners2018b} and HD~180617~b \citep{Kaminski2018} orbit their host stars within the habitable zone. \cite{Nagel2019a} also presented a Neptune-mass planet with high eccentricity. 
Other planetary systems have been detected by \cite{Trifonov2018}, \cite{Sarkis2018}, \cite{Ribas2018}, \cite{Luque2018}, \cite{Zechmeister2019}.

To be able to characterize a planetary system it is important to determine fundamental stellar parameters such as the stellar mass and radius, effective temperature $T_{\rm eff}$, surface gravity $\log{g}$, 
metallicity, and luminosity. Different ways to determine the first two properties are discussed in \citet[][hereafter Schw19]{Schweitzer2019}. 
One approach to deriving the photospheric parameters $T_{\rm eff}$, $\log{g}$, and [Fe/H] is the analysis of stellar spectra. 
M-dwarf spectra are more complex than those of Sun-like stars due to the lower temperatures in M-dwarf atmospheres. Molecular lines produce forests of spectral features and make the determination 
of atmospheric parameters more difficult. This requires a full spectral synthesis instead of analyzing and modeling individual lines independently of the underlying atmosphere as it is done for Sun-like 
stars (MOOG -- \citet{Sneden1973,Sneden2012}, SME -- \citet{Valenti2012}). 

Most recent generations of stellar atmosphere models are capable of accurately reproducing the spectral features present in cool star spectra. The PHOENIX\ code, developed by 
\cite{Hauschildt1992,Hauschildt1993}, is one of the most advanced stellar 
atmosphere codes. This code takes into account molecule formation in cool stellar atmospheres and is, therefore, especially suited to model M-dwarf atmospheres. 
The code was updated several times and the latest grid of stellar atmospheres was published by \cite{Husser2013}. 

Fundamental stellar parameters of low-mass M dwarfs have been determined with different methods in different wavelength ranges throughout the literature. 
\citet[][hereafter GM14,]{GaidosMann2014} observed low-resolution spectra of 121 M dwarfs in the near-infrared $JHK$ bands and around half of them in the visible range using SpeX, and the SuperNova 
Integral Field Spectrograph (SNIFS), respectively. 
To determine effective temperatures for stars with NIR spectra, they calculated spectral curvature indices from the $K$-band. For stars with VIS spectra they fit BT-Settl models \citep{Allard2011}, 
which were calculated from the PHOENIX code and describe atmospheres of cool M, L, and T dwarfs. Both the VIS and NIR ranges were used to derive metallicities from relations of atomic line strength as 
described in \cite{Mann2013}. The same relations were also used by \cite{Rodriguez2018} to determine metallicities from mid-resolution $K$-band spectra for 35 M dwarfs of the K2 mission. 
The NIR $K$-band was also investigated by \citet[][hereafter RA12,]{RojasAyala2012} to determine effective temperatures and metallicities of 133 M dwarfs from low-resolution TripleSpec spectra ($R \sim$ 2\,700). 
They calculated the H$_2$O-K2 index to quantify the absorption from H$_2$O opacity and derive effective temperatures. The calibration was done using BT-Settl models \citep{Allard2012} of solar 
metallicity. \cite{Newton2014} derived metallicity relations based on equivalent widths using $JHK$-band low-resolution spectra ($R \sim$ 2\,000), calibrated with multiple systems containing at least an 
M-dwarf secondary and a main-sequence primary of spectral type F, G, or K. 
\cite{Birky2017} used PHOENIX models for modeling the stellar parameters $T_{\rm eff}$, $\log{g}$, and metallicity for late-M and early-L dwarfs from high-resolution, near-infrared SDSS/APOGEE spectra 
($R \sim$ 22\,500). 
High-resolution APOGEE spectra have also been used by \cite{Souto2017}, who fit MARCS models to determine abundances for thirteen elements of the exoplanet-host M dwarfs Kepler-138 and 
Kepler-186. \cite{Souto2018} derived $T_{\rm eff}$, $\log{g}$, and chemical abundances of eight elements of the exoplanet-host Ross 128 (M4.0\,V) by fitting MARCS and BT-Settl models \citep{Allard2013}. 
Using the near-infrared $Y$-band \cite{Veyette2017} derived precise effective temperatures as well as Ti and Fe abundances from high-resolution spectra of 29 M dwarfs by combining spectral synthesis, empirical 
calibrations, and equivalent widths. With the same method, using BT-Settl models, \cite{Veyette2018} determined $T_{\rm eff}$, $\log{g}$, and [Ti/Fe] of eleven planet-host M dwarfs from CARMENES 
$Y$-band spectra. However, they convolved the CARMENES spectra to match a resolution of 25\,000 instead of the original 80\,500, which led to a loss of spectral information. 
\citet[hereafter Pass18]{Passegger2018} derived the parameters $T_{\rm eff}$, $\log{g}$, and [Fe/H] for the CARMENES sample from fitting PHOENIX-ACES \citep{Husser2013} models 
to high-resolution CARMENES spectra in the VIS. More recently, Schw19 used the same method to provide updated parameters for this sample together with stellar mass and radius.
A work by \citet[][hereafter Raj18]{Rajpurohit2018} combined the VIS and the NIR wavelength ranges of the publicly available CARMENES spectra of \cite{Reiners2018a} to determine stellar parameters from 
BT-Settl model fits. 

A widely neglected property in spectroscopic parameter determination so far is the stellar magnetic field. In M dwarfs, the stellar magnetic field plays an important role as a driver for 
activity. A typical, averaged surface strength of these fields, $B_s$, is about 1--2\,kG in the majority of active M dwarfs, but can go well beyond 4~kG in some of them 
\citep{Shulyak2017}. Generally, the magnetic field affects the shapes of spectral lines according to their magnetic sensitivity, described by the Land\'e g-factors, and the number and strength 
of individual Zeeman components \citep[so-called Zeeman pattern,][]{Landi2004}. Due to a finite spectral resolution and non-zero stellar rotation, the individual Zeeman components are normally 
not resolved and the magnetic field manifests itself as a Zeeman broadening. The magnitude of the Zeeman broadening scales as $\Delta\lambda \varpropto g_{\rm eff}\lambda_{\rm 0}^2 B_s$, where $g_{\rm eff}$ 
is the effective Land\'e-factor, $\lambda_{\rm 0}$ is the central wavelength of the line, and $B_s$ is the strength of the magnetic field. Given the quadratic dependence on the wavelength, the effect 
of Zeeman broadening can be comparable to or even stronger than the rotation broadening in slow and moderately rotating M dwarfs ($v\sin{i}$ < 10 km\,s$^{-1}$) in the NIR wavelength domain, in contrast to the 
lines in the VIS spectral range. In addition, when the stellar rotation is very fast ($v\sin{i}$ > 10 km\,s$^{-1}$), and the rotational broadening is the dominant broadening mechanism, the magnetic 
field can still affect the line depths and corresponding equivalent widths via magnetic intensification \citep{Landi2004,Shulyak2017}. As fast rotating stars always host strong magnetic fields 
according to the rotation-activity relation \citep{Reiners2009}, the effect of the magnetic field cannot be fully ignored in the analysis of spectral lines even in these stars. Therefore, additional 
care should be taken in the analysis to exclude lines that demonstrate high magnetic sensitivity. 

So far, stellar parameters have been determined either from the VIS or the NIR. Although Raj18 combined both wavelength regimes, no direct comparison between parameters derived 
separately from the VIS and from the NIR has been made. In this work, we analyze high signal-to-noise ratio (S/N), high-resolution CARMENES spectra in the VIS, NIR, and VIS+NIR ranges of 342 M dwarfs. We 
fit an updated version of the PHOENIX-ACES models \citep{Husser2013}, the so-called PHOENIX-SESAM models, to derive $T_{\rm eff}$, $\log{g}$, and [Fe/H]. We compare results for the CARMENES sample stars 
derived from different wavelength ranges with each other and with literature values and discuss our findings.


\section{Observations}
\label{Observations}

We observed 342 stars of spectral types between M0.0\,V and M9.0\,V with CARMENES in the VIS and NIR channels. These observations were obtained between January 2016 and January 2019. 

For wavelength calibration, the CARMENES VIS spectrograph is equipped with U-Ne, U-Ar, and Th-Ne hollow-cathode lamps, and the NIR spectrograph with U-Ne hollow-cathode lamps. In addition, a 
temperature- and pressure-stabilized Fabry-P\'{e}rot etalon \citep{Schaefer2018} is used as a calibration source and to measure nightly drifts of the spectrographs \citep{Bauer2015}. 
The spectrum extraction is performed automatically by the CARMENES data reduction pipeline CARACAL \citep{Zechmeister2014,Caballero2016}. The process is based on the REDUCE package from 
\cite{PiskunovValenti2002} 
and includes bias subtraction, flat fielding, and cosmic ray detection. For the CARMENES planet search survey, radial velocities are derived using the radial velocity pipeline SERVAL 
\citep[SpEctrum Radial Velocity AnaLyser;][]{Zechmeister2018}. The code corrects each spectrum for barycentric motion \citep{WrightEastman2014} and secular acceleration \citep{Zechmeister2009} 
before co-adding them to construct a high-S/N template spectrum of every target star. The radial velocities of each single spectrum are then computed by least-squares fitting against the template 
spectrum. 
A template spectrum is generated for each target star from at least five individual spectra. Due to the high S/N, we used these template spectra to determine stellar parameters. 
As Pass18 showed, spectra with S/N < 75 can lead to unrealistically high or low temperatures and metallicities during the fitting process due to bad quality spectra. However, not all 
template spectra satisfy this criterion of S/N > 75, which is why we excluded 34 templates from the VIS channel, 15 templates from the NIR channel, and 18 combined templates in VIS+NIR of the 
342 stars from our sample. 
We also excluded nine double-line spectroscopic binaries found in the CARMENES data by \cite{Baroch2018}.

The telluric contamination in the spectral range that we investigated with the VIS channel is mainly dominated by O$_2$ and H$_2$O absorption bands \citep{Passegger2017}. However, the 
contamination is concentrated around the 
K~{\sc i} doublet at around 768~nm and the Na~{\sc i} doublet at around 819~nm, where we used masks to exclude telluric lines. Other contributions of telluric lines are negligible in the VIS 
wavelength range. 
For the NIR the situation is different as strong bands of telluric features contaminate almost the entire stellar spectrum. A common method for telluric correction is the observation of a hot star
with few and broad stellar lines. If observations of a telluric standard are not possible, because of a lack of suitable stars in the observed region on the sky or time constraints, the telluric 
spectrum can also be modeled and subtracted from the observed one. We used the telluric-correction tool Molecfit \citep{Kausch2014,Smette2015} to model the atmospheric absorption of individual 
molecules. The code incorporates a radiative transfer model together with the high-resolution transmission molecular absorption database, HITRAN \citep{Gordon2017}, and atmospheric profiles to calculate 
synthetic transmission spectra. 
Using a Levenberg-Marquardt algorithm the transmission model is then adjusted to match the molecular column densities of the atmospheric constituents. The result is the observed spectrum corrected with 
the best-fit telluric model. We used these telluric-corrected NIR-channel spectra to calculate telluric-free high-S/N templates with SERVAL. Further details on the telluric absorption 
correction will be provided in a forthcoming publication of the CARMENES series.

\section{Method}
\label{Method}

We followed the method described in Pass18. In that study, we derived the effective temperature $T_{\rm eff}$, surface gravity $\log{g}$, and metallicity [Fe/H] for 
300 stars from high-resolution spectra in the visible wavelength range by fitting the latest PHOENIX-ACES model grid presented by \cite{Husser2013} to the observed spectra using a two-step 
procedure described in the following. We used the iron abundance [Fe/H] that is closely related to the metal abundance [M/H], which is a proxy for metallicity $Z$. In the 
first step a coarsely spaced model grid was explored around the expected parameters of the target star. The $\chi^2$ was calculated to get a rough global minimum, which served as starting point for 
the second step. There the model grid was linearly interpolated to explore the global minimum on a finer grid. We analyzed the quadratic interpolation of the model grid, but found no improvement compared to 
linear interpolation. Hence, we linearly interpolated our model grid to save computation time. To calculate the $\chi^2$ from the observed spectrum and the model, the model spectra were 
convolved with a Gaussian function to adapt them to the instrumental resolution. The wavelength grid of the models was linearly interpolated to match the wavelength grid of the observed spectrum. 
Both, the average observed flux and the model flux were normalized to unity using a linear fit to the pseudo-continuum. We included rotational broadening for different $v\sin{i}$ values of 
our sample stars, taken from \cite{Reiners2018a}. To do so we used a broadening function that estimates the effect on the line-spread function due to stellar rotation. The model spectrum was 
then convolved with the resulting line-spread function. 

\begin{figure*}[htb]
 \includegraphics[width=0.95\textwidth]{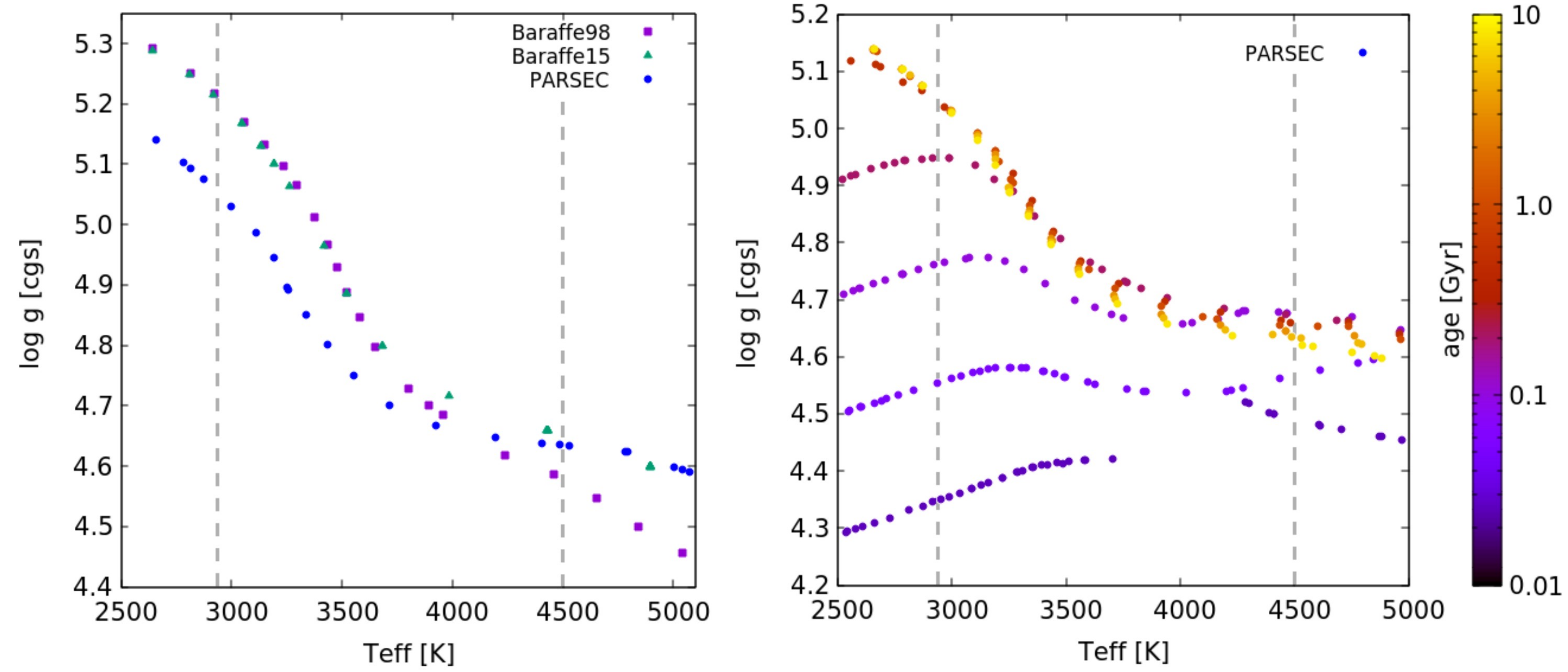}
 \caption{Surface gravity $\log{g}$ as a function of effective temperature $T_{\rm eff}$. The left panel compares the Lyon group's BCAH98 \citep{Baraffe1998}, BHAC15 \citep{Baraffe2015} models, 
 and PARSEC models for [M/H] = 0 and fixed age of 5\,Gyr. The right panel shows the PARSEC models for [M/H] = 0 and variable age from 0.01 to 10\,Gyr. The gray vertical lines indicate the 
 temperature range of our PHOENIX-SESAM grid from 2900\,K to 4500\,K. }
 \label{fig:evol_models}
\end{figure*}

As suggested in Pass18 we determined $\log{g}$ from evolutionary models to break degeneracies between the parameters. In contrast to that work, we used evolutionary models from the PARSEC v1.2S library 
\citep{Bressan2012,Chen2014,Chen2015,Tang2014}, which provide, among other parameters, $T_{\rm eff}$ and $\log{g}$ for different stellar ages and metallicities in the range $-$2.2 < [M/H] < +0.7. 
In contrast, the models from \cite{Baraffe2015} provided only solar-metallicity isochrones. The large range of metallicities provided by the PARSEC models avoids extrapolation beyond +0.7~dex in 
most cases. Also, the finer sampling rate of metallicities reduces the error from interpolation, which is why the PARSEC models were preferred for this work. Figure~\ref{fig:evol_models} shows a comparison 
between the \cite{Baraffe1998} evolutionary models used in Pass18, the updated version from \cite{Baraffe2015}, and the PARSEC models (left panel), as well as the PARSEC models for different ages 
(right panel). 

Additionally, we included age estimates for our target stars. \cite{Cortes2016} gathered proper motions from the literature or computed them where not available, and calculated Galactic
space velocities for all Carmencita target stars with radial velocity measurements in order to kinematically identify their membership in young moving groups, and thin and thick disk populations as in \cite{Montes2001}. 
Following their method, we updated kinematics of the target stars with the latest {\em Gaia} DR2 proper motions and parallaxes \citep{GAIA2016,GAIA2018}. The results of this study will be published 
in a subsequent paper.

From this, we used the mean age for each of these kinematic populations and associations. As most of our sample are stars not belonging to the young disk, we assumed an average age of 5\,Gyr. 
For stars belonging to the young disk or young moving groups with ages younger than 1\,Gyr we used the corresponding PARSEC models at young ages. 
Based on $T_{\rm eff}$ and metallicity chosen by our algorithm, $\log{g}$ was calculated from the $T_{\rm eff}$-$\log{g}$ relations. As described in Pass18, 
the relations were linearly interpolated for metallicities between $-$1.0 and $+$0.7. With these three parameters we interpolated the corresponding PHOENIX models and calculated the $\chi^2$. 
We applied this procedure with an updated PHOENIX model grid to high-resolution and high-S/N CARMENES template spectra in the VIS and the NIR, as well as the VIS+NIR combined. A short description of the new 
PHOENIX grid follows in Section~\ref{Models}. 

We investigated a method for reducing the dependency on evolutionary models. For this, we determined the stellar parameters for a subsample of stars as described above. Then we followed the approach described 
by Schw19 by calculating the radius from $T_{\rm eff}$ and the total luminosity using the Stefan-Boltzmann law and the mass from a linear mass-radius relation (Schw19). From mass and radius we derived 
$\log{g}$ and inserted this value as a fixed parameter into 
our algorithm, which left two free parameters. With this we determined $T_{\rm eff}$ and metallicity. The procedure can be repeated iteratively until the parameters converge. However, after the 
first iteration the differences between the parameters were not significant. For this reason we followed the procedure mentioned before. 

We analyzed the effects on the resulting stellar parameters when using a finer model grid from which we can interpolate. It also served to test if linear interpolation of the standard model grid is 
acceptable. Hence, we calculated model atmospheres with finer grid spacing in the framework of the grid published by \cite{Husser2013} (hereafter referred to as standard step-size grid). The finer grid 
ranges from 
2800\, -- 4300\,K in steps of 50\,K (instead of 100\,K), from 3.0 to 6.0\,dex in $\log{g}$ and from --1.0 to +1.0 in [Fe/H] with a step size of 0.2 (instead of 0.5). This grid was used in our algorithm as 
described 
above to derive the stellar parameters from interpolation between the grid points for a subsample of 100 stars. We find no significant difference between the parameters calculated from the finer grid 
and the parameters from the standard step-size grid. The maximum deviations are still smaller than the estimated errors of the fitting procedure (see Table~\ref{tab:errors}). Thus, we conclude that 
for our purpose linear interpolation of the standard step-size grid is sufficient and a finer model grid is not required. 


 \begin{center}
   \begin{table}[htb]
     \caption{Uncertainties for stellar parameters for different wavelength ranges and $v\sin{i}$.}
     \label{tab:errors}
     \centering %
     \begin{tabular}{clccc}
       \hline \hline 
             
        $v\sin{i}$ & Parameter & VIS+NIR & NIR & VIS\\
           \hline 
       $\sim$ 2 km\,s$^{-1}$ & $\Delta T_{\rm eff}$ [K] & 54 & 56 & 51\\
                      & $\Delta \log{g}$ [dex] & 0.06 & 0.04 & 0.04\\
                      & $\Delta$ [Fe/H] [dex] & 0.19 & 0.16 & 0.16\\ \\
             
      2--5 km\,s$^{-1}$ & $\Delta T_{\rm eff}$ [K] & 64 & 72 & 64\\
                      & $\Delta \log{g}$ [dex] & 0.07 & 0.05 & 0.05\\
                      & $\Delta$ [Fe/H] [dex] & 0.19 & 0.16 & 0.17\\ \\
              
      5--10 km\,s$^{-1}$ & $\Delta T_{\rm eff}$ [K] & 100 & 124 & 85\\
                      & $\Delta \log{g}$ [dex] & 0.08 & 0.07 & 0.07\\
                      & $\Delta$ [Fe/H] [dex] & 0.23 & 0.23 & 0.21\\ \\
              
      10--15 km\,s$^{-1}$ & $\Delta T_{\rm eff}$ [K] & 131 & 162 & 108\\
                      & $\Delta \log{g}$ [dex] & 0.10 & 0.10 & 0.09\\
                      & $\Delta$ [Fe/H] [dex] & 0.29 & 0.33 & 0.28\\ \\
              
      15--20 km\,s$^{-1}$ & $\Delta T_{\rm eff}$ [K] & 134 & 162 & 136\\
                      & $\Delta \log{g}$ [dex] & 0.11 & 0.11 & 0.12\\
                      & $\Delta$ [Fe/H] [dex] & 0.33 & 0.38 & 0.38\\ \\
              
      > 20 km\,s$^{-1}$ & $\Delta T_{\rm eff}$ [K] & 124 & 170 & 162\\
                      & $\Delta \log{g}$ [dex] & 0.12 & 0.11 & 0.13\\
                      & $\Delta$ [Fe/H] [dex] & 0.40 & 0.46 & 0.39\\
 
          \hline
     
       \hline
     \end{tabular}
   \end{table}
 \end{center}
 
For error estimation, we applied the method from \cite{Passegger2016}. We generated 1400 model spectra with randomly distributed parameters and added Poisson noise to simulate S/N $\sim$ 100. These spectra 
served as input for our algorithm with which we recovered the input parameters. The errors were determined as the standard deviations from the mean value in the residual parameter distributions. For 
this work, we also calculated errors using different rotational velocities, since this value can have a large influence on the derived stellar parameters. Our derived uncertainties for the VIS, NIR, and 
VIS+NIR combined are presented in Table~\ref{tab:errors}. For $\log{g}$ the uncertainty is mainly dominated by the fitting 
procedure, given that the error coming from interpolation or extrapolation of the $T_{\rm eff}$-$\log{g}$ relations is smaller than 0.05~dex and therefore negligible. 

\subsection{Models}
\label{Models}

 \begin{figure*}[htb]
  \includegraphics[width=0.99\textwidth]{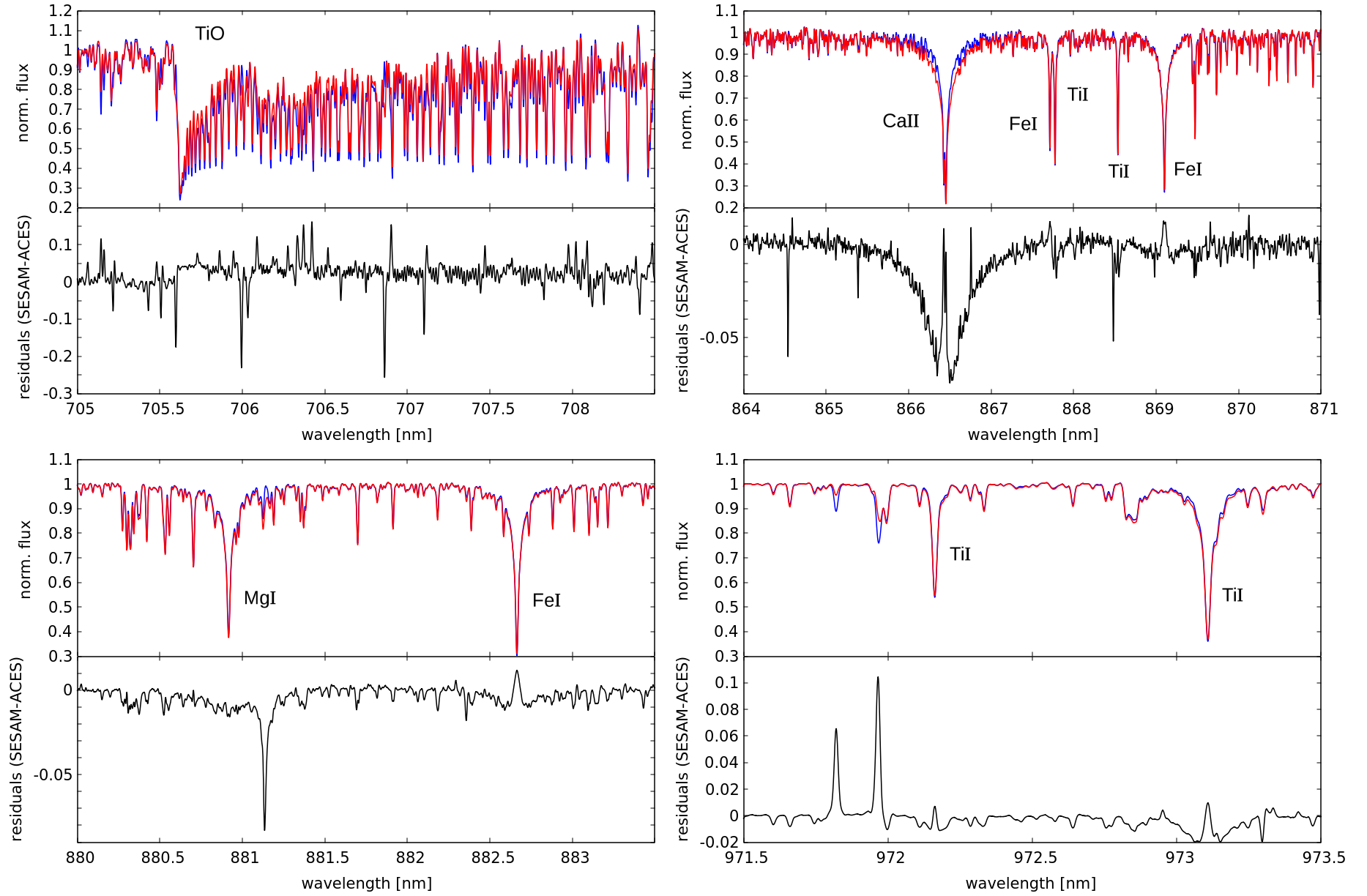}
  \caption{Comparison of PHOENIX-ACES (blue) and PHOENIX-SESAM (red) models for a selection of lines. The parameters of the models are $T_{\rm eff}$=3500~K, $\log{g}$=5.0~dex, and [Fe/H]=0.0~dex. The models 
    are broadened corresponding to the spectral resolution in the VIS ($R \approx$ 94\,600) and NIR ($R \approx$ 80\,500) channel. For each wavelength region the residual between the two models are shown.}
  \label{fig:sesam-aces}
  \end{figure*}

The PHOENIX atmosphere models we used for our analysis are based on the PHOENIX code developed by \cite{Hauschildt1992,Hauschildt1993}. The code has undergone continuous improvement as shown in 
\citet{Hauschildt1997}, \citet{HauschildtBaron1999}, \citet{Claret2012}, \citet{Husser2013}. It computes one-dimensional (1D) model atmospheres for plane-parallel and spherically symmetric stars such as 
main sequence stars and brown dwarfs down to L and T spectral types; and white dwarfs as well as giants, including accretion disks and expanding envelopes of novae and supernovae. It can be executed in 
radiative transfer mode in local or non-local thermodynamic equilibrium, and synthetic spectra can be calculated in 1D or 3D. 
The PHOENIX code serves as a basis for several model grids of cool stars, for example, the NextGen models \citep{Hauschildt1999}; the AMES models \citep{Allard2001} accounting for condensation and 
dust formation in two different approximations (AMES-dusty and AMES-cond); the BT-Settl models \citep{Allard2011} using yet another dust approximation for very low temperature atmospheres down to the 
planetary mass regimes; and the PHOENIX-ACES models \citep{Husser2013} using an updated equation of state for cool dwarfs. This last grid used a new equation of state, especially designed for molecule 
formation inside the coolest known stellar atmospheres. 

Most recently, we calculated a new grid for temperature range between 2900\,K and 4500\,K following \cite{Husser2013} using our latest equation of state SESAM \citep{Meyer2017}. We used solar 
abundances as reported by \cite{Asplund2009}, updated with values from \cite{Caffau2011}, as well as updated atomic and molecular line lists. This new PHOENIX-SESAM grid was incorporated in our procedure 
for parameter determination. 
An extended grid including a detailed description will be presented in a subsequent paper. Figure~\ref{fig:sesam-aces} presents a comparison between the PHOENIX-ACES (blue) and PHOENIX-SESAM (red) 
models for a selection of lines. Some differences can be seen in the TiO-band ($\lambda$ > 705~nm) and Ca~{\sc ii} line ($\lambda$ 866.45\,nm), whereas other Ti- and Fe-lines are not significantly 
influenced. 

\subsection{Line selection}

 \begin{center}
   \begin{table}
     \caption{Wavelength regions and lines used for parameter determination.}
     \label{tab:lines}
     \centering %
     \begin{tabular}{lcc}
       \hline \hline 
             
        Line & $\lambda$ [nm] & Channel\\
           \hline 
        $\gamma$-TiO & > 705.5 & VIS\\
        Ca~{\sc i} & 715.0 & VIS\\
        K~{\sc i} & 770.1 & VIS \\
         & 1177.3, 1177.6, 1252.55, & NIR\\ 
         & 1516.7, 1517.2 & NIR\\
        Ti~{\sc i} & 841.5, 842.9, 843.7, 843.8,  & VIS \\
         & 846.9, 867.8, 868.5 & VIS \\
         & 972.15, 973.1, 983.5, & NIR\\
         & 1058.75, 1066.45, 1077.8 & NIR \\
        Fe~{\sc i} & 847.1, 851.6, 867.7, & VIS \\
         & 869.1, 882.7 & VIS \\
         & 1112.3 & NIR\\
        Mg~{\sc i} & 880.9 & VIS\\

          \hline
     
       \hline
     \end{tabular}
   \end{table}
 \end{center}

\begin{figure*}[htb]
 \includegraphics[width=0.95\textwidth]{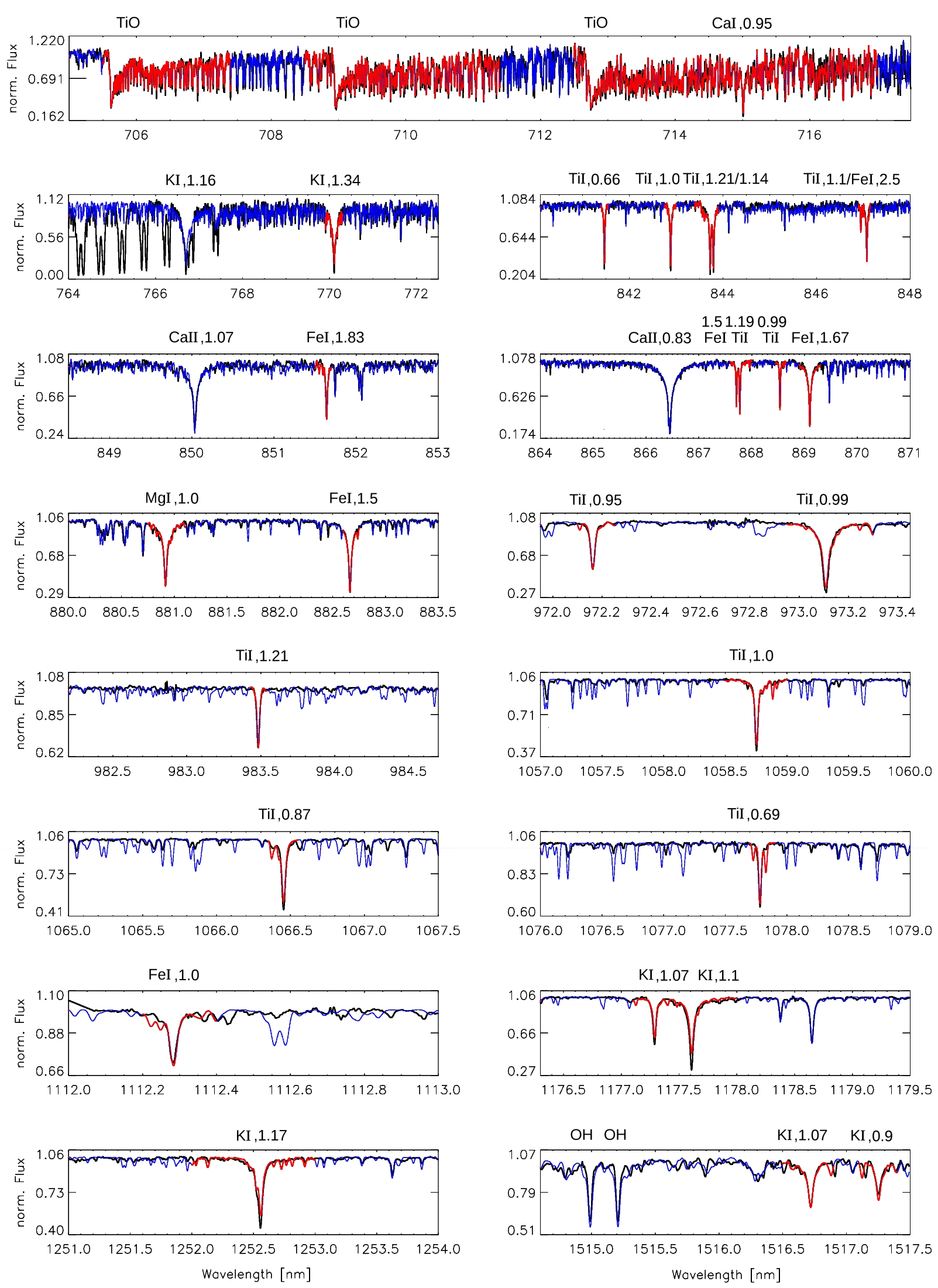}
 \caption{CARMENES template spectrum of GJ 133 = J03213+799 (black) with best-fit model (blue and red). The fitting regions are marked in red. All lines are identified together with their Land\'e factors when 
 available. }
 \label{fig:best-fit}
 \end{figure*}

An appropriate set of spectral lines is crucial for the accurate determination of stellar parameters. Unlike studies by Raj18, who used all strong lines available in the spectra, we 
carefully selected the lines that we used by the following criteria. We investigated the sensitivity in $T_{\rm eff}$ and metallicity of several atomic and molecular lines in the NIR by calculating the 
deviation between models with different parameters. For this purpose, we 
selected reference model spectra with $T_{\rm eff}$ between 2800\,K and 4000\,K, [Fe/H] between --1.0 and +1.0, and with $\log{g}$ fixed to 5.0 for simplicity, and because we constrain this parameter from the 
PARSEC evolutionary models. Each reference spectrum was compared to all model spectra of the selected grid, and we calculated the deviations for each line. Then, we selected lines that showed high 
deviations, which is high sensitivity over a wide parameter range. As discussed in Pass18 we excluded the Ca~{\sc ii} lines at 850.05\,nm and 866.45\,nm due to possible chromospheric emission, and the Na~{\sc i} 
doublet around 819\,nm because of degeneracies in the strength and width of the lines. We also excluded some atomic and molecular lines for which the models showed shortcomings in accurately fitting the 
line shape. 

As described above, the telluric lines in the CARMENES spectra are corrected by modeling atmospheric transmission spectra. This method, like others, suffers from imperfections due to inaccurate weather 
data or uncertainties in the atmospheric molecular line list. To avoid these imperfections we chose spectral lines that are not heavily affected by telluric features. 
Finally, yet importantly, we specifically excluded atomic lines with large Land\'e-factors ($g_{\rm eff} > 1.5$) in the NIR and lines that are blended with lines having large Land\'e-factors. 
The factors were computed using transition parameters extracted from the current edition of the Vienna Atomic Line Database \citep[VALD,][]{Ryabchikova2015} or using the LS coupling 
approximation when this information was not available. In this way we 
minimized (but not completely exclude) the impact of the magnetic field on our determination of atmospheric parameters. As shown before, the Zeeman broadening scales with the square of the wavelength, 
so although lines in the VIS are not noticeably affected, lines with the same Land\'e-factors in the NIR can be broader, which might have an impact on the parameter determination. 
The lines of molecular species in NIR are generally very weakly magnetic sensitive or not magnetic sensitive at all, and many of them could be safely used in the analysis of atmospheric properties provided that their 
transition parameters are accurately known. Unfortunately, this is often not the case, as for OH ($\lambda$ > 1510\,nm) where we found that the models fit only poorly. The only exception is the 
Wing-Ford band of FeH lines around 980\,nm, whose transition parameters are well known but their magnetic sensitivity is extremely high \citep{Berdyugina2002,Shulyak2014}, which limits using these 
lines in spectroscopic parameter determinations. On these grounds, we found only a few suitable lines, some of them in the NIR range, which makes them interesting for parameter determination. 

Selecting spectral lines according to their magnetic sensitivity makes our analysis of NIR spectra superior to other similar works. We expect that at least one third of our sample stars have kG 
level magnetic fields according to their activity indicators \citep{Schoefer2019} and we aim at minimizing possible biases in the derived stellar parameters. In Fig.~\ref{fig:best-fit} all lines that we 
used in the VIS and NIR are identified together with their Land\'e factors when available. Table~\ref{tab:lines} summarizes the 28 lines and molecular bands we used for parameter determination.
\section{Results and discussion}

We visually inspected the best fits for all stars in our sample to ensure a good fit quality and reliable stellar parameters. During this process we removed more stars from our final parameter list, 
because strong stellar activity or high rotational velocity led to an insufficient model fit to the spectra. Due to the fact that magnetic broadening has a larger effect on lines in the NIR range, as 
discussed before, 
we excluded more stars with parameters derived from the NIR range alone. This results in final stellar parameters derived in the VIS for 275 M dwarfs, in the NIR for 271 M dwarfs, and in VIS+NIR for 
276 M dwarfs. The final sample of 282 M dwarfs is presented in Table~\ref{tab:sample} showing the name, coordinates ({\em Gaia} DR2), spectral type 
\citep[Carmencita,][]{Caballero2016a}, assumed age according to their kinematics, $v\sin{i}$ \citep[][if not indicated otherwise]{Reiners2018a}, and an activity flag. The derived parameters 
are listed in Table~\ref{tab:results} for 
the different wavelength ranges. In the following we compare our results derived from multiple wavelength ranges, as well as to values from the literature, and discuss outliers. 

\subsection{Comparison of different wavelength ranges}

\begin{figure*}[htb]
\centering
 \includegraphics[height=0.95\textheight]{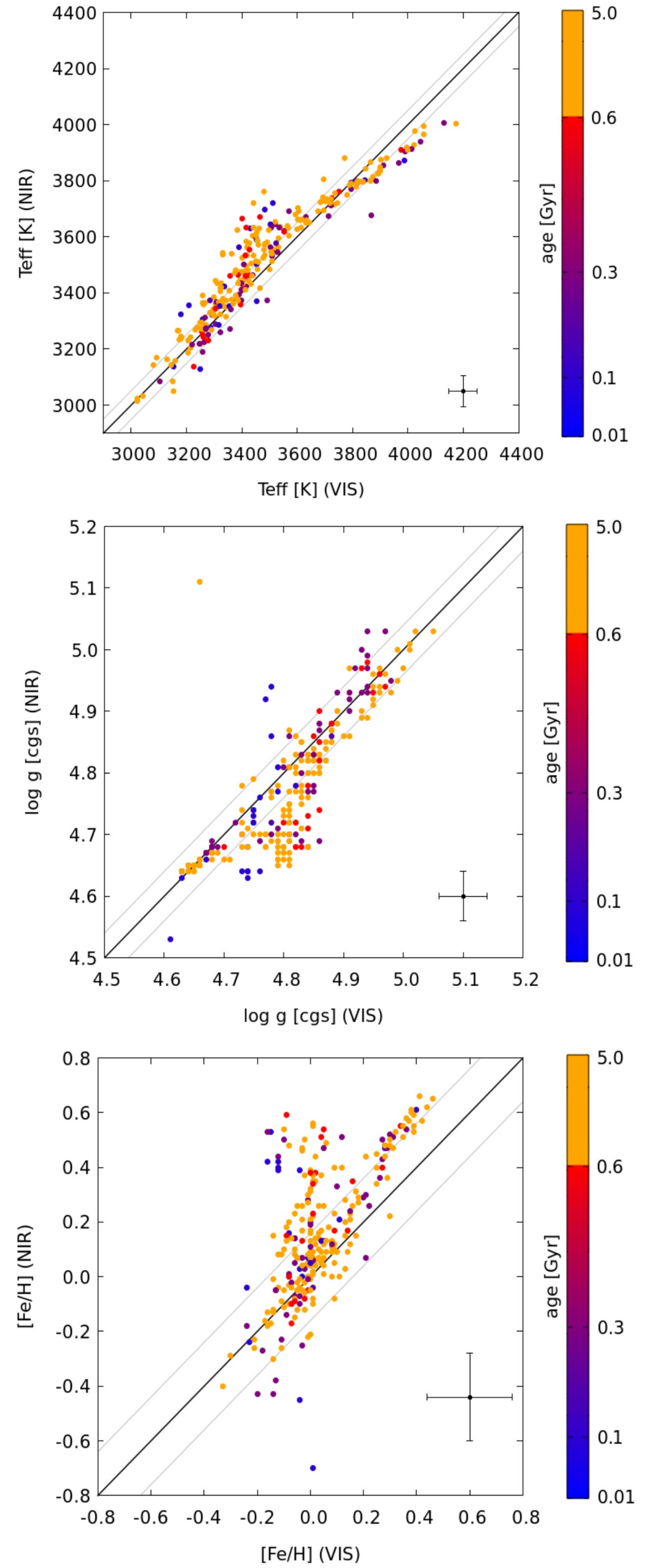}
 \caption{Comparison of stellar parameters derived from VIS and NIR ranges. The assumed stellar age is color-coded. The black line indicates the 1:1 relation, the gray lines represent the 
 1$\sigma$ deviations for VIS of 51~K, 0.04~$\log{g}$, and 0.16~[Fe/H]. Error bars for VIS and NIR are plotted in the lower right corner of each plot. }
 \label{fig:comparison}
 \end{figure*}
 
Figure~\ref{fig:comparison} shows comparison plots for the different parameters. The upper panel presents $T_{\rm eff}$, with the stellar age color-coded. In general, the temperatures derived 
from the VIS and NIR wavelength ranges follow the 1:1 relation within their errors. A group of outliers is located above the 1:1 relation between 3300\,K and 3600\,K, showing slightly higher temperatures 
compared to the VIS. For temperatures higher than \textasciitilde3800\,K values derived from the NIR seem to be about 100--200\,K lower. However, temperatures determined from VIS+NIR for the same stars 
almost perfectly correspond to their values in the VIS. This is shown in Fig.~\ref{fig:comparison_VIS+NIR} in the Appendix. 

Also $\log{g}$ (middle panel) corresponds very well in the two wavelength ranges. Young stars with ages less than 0.1\,Gyr are mainly located at the lower end. Most of them have ages less than 50~Myr and are 
still contracting \citep[e.g.,][]{Palla2002}, which explains their lower $\log{g}$. We find some trend toward lower $\log{g}$ values for the NIR range, where we see a group with values about 0.1 -- 0.2\,dex 
lower. They are related to the outlier group found in temperature, since our determination of $\log{g}$ depends on $T_{\rm eff}$ and [Fe/H]. A clear outlier can be seen at $\log{g}$ 5.11\,dex in the NIR. 
This value is about 0.5\,dex higher compared to $\log{g}$ derived from the VIS and VIS+NIR, and belongs to J21152+257. The star also exhibits a 100\,K higher temperature and 0.3\,dex higher 
metallicity in the NIR. The deviation of its parameters might be explained by the model fit being worse in the NIR range. There are no literature values for this star, but the spectral type of M3.0\,V agrees 
better with the temperatures derived in VIS and VIS+NIR. 

The bottom panel of Fig.~\ref{fig:comparison} shows the comparison in metallicity. 
While most stars lie on the 1:1 relation within their errorbars, there is a large group with values larger in the NIR compared to the VIS. This corresponds to what is shown in 
Fig.~\ref{fig:lit_NIR} and Fig.~\ref{fig:lit_VISNIR}, where we find most of our sample to be slightly more metal-rich compared to literature. 

\subsection{Literature comparison}
\begin{figure*}[htb]
\centering
 \includegraphics[height=0.95\textheight]{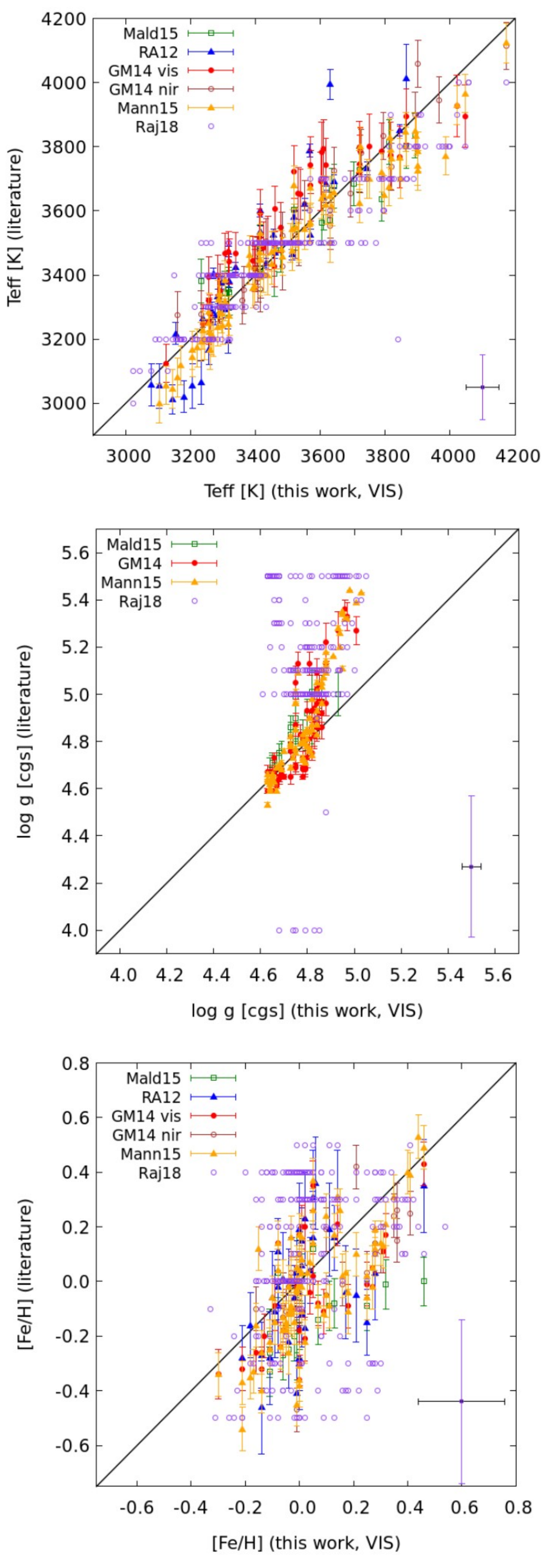}
 \caption{Comparison between results from VIS and literature values for $T_{\rm eff}$ (top panel), $\log{g}$ (middle panel), and [Fe/H] (bottom panel). The 1:1 relation is 
 indicated by the black line. The 
 uncertainties of this work (black) are shown in the lower right corner of each panel together with the uncertainties of \cite{Rajpurohit2018} (purple).}
 \label{fig:lit_VIS}
 \end{figure*}

\begin{figure*}[htb]
\centering
 \includegraphics[height=0.95\textheight]{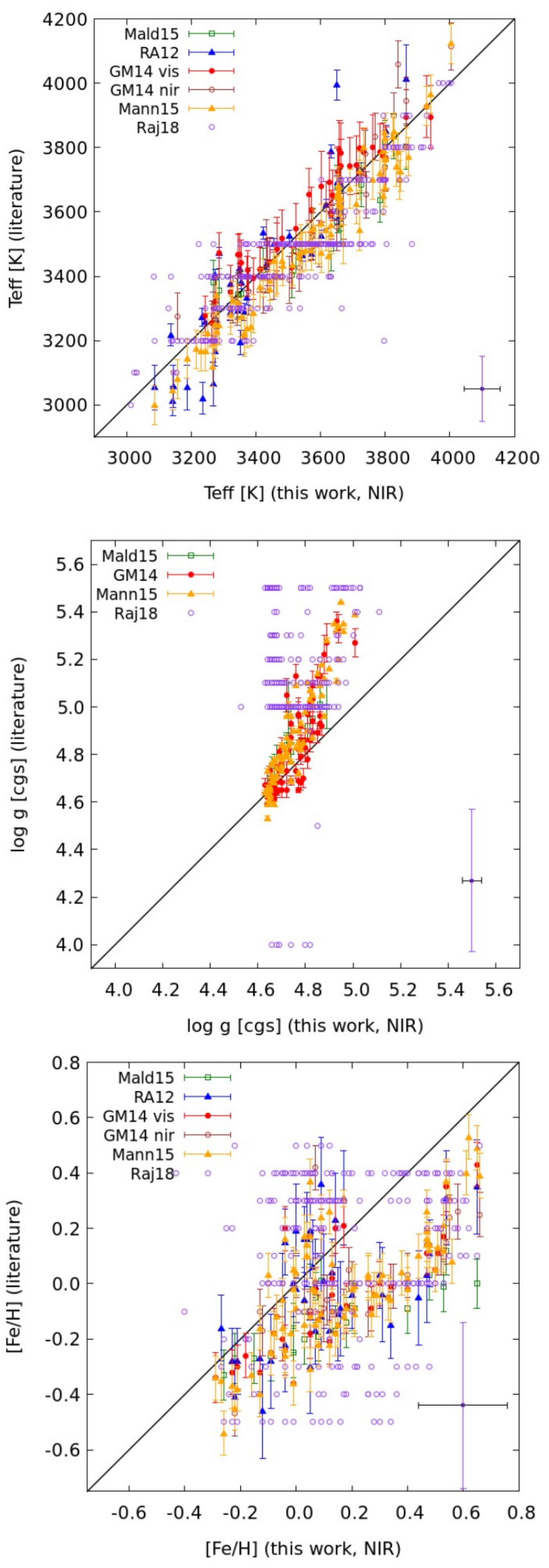}
 \caption{Comparison between results from NIR and literature values for $T_{\rm eff}$ (top panel), $\log{g}$ (middle panel), and [Fe/H] (bottom panel). The 1:1 relation is 
 indicated by the black line. The 
 uncertainties of this work (black) are shown in the lower right corner of each panel together with the uncertainties of \cite{Rajpurohit2018} (purple).}
 \label{fig:lit_NIR}
 \end{figure*}

\begin{figure*}[htb]
\centering
 \includegraphics[height=0.95\textheight]{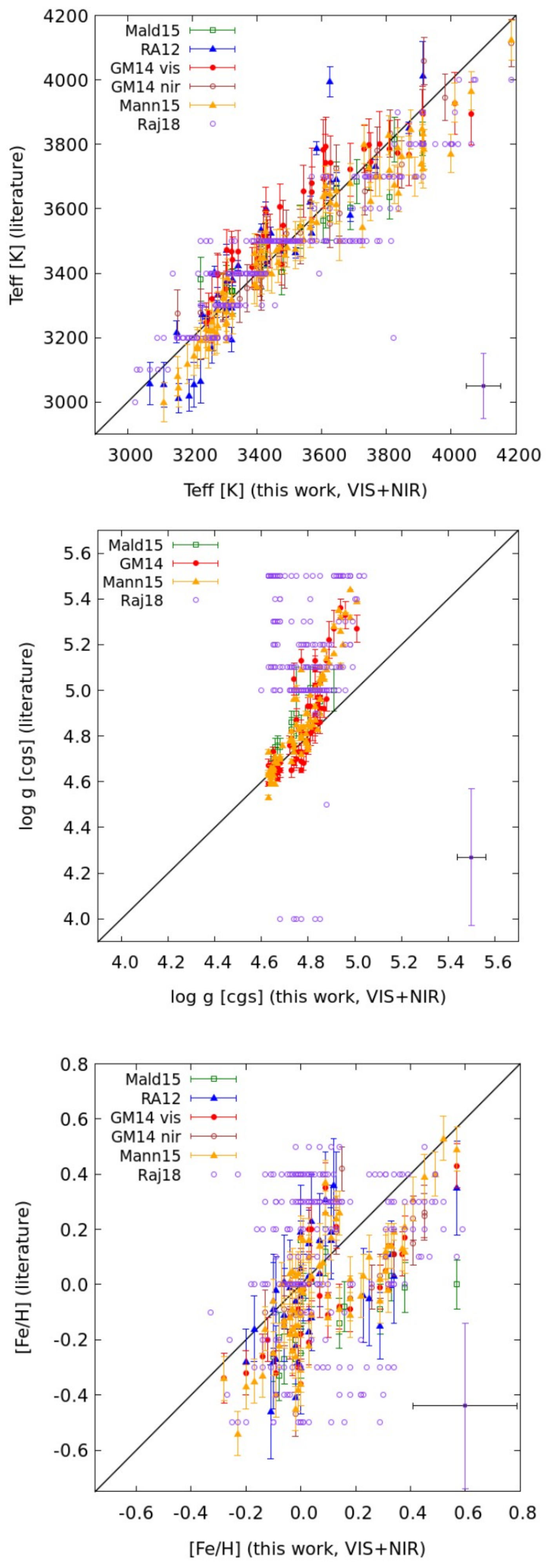}
 \caption{Comparison between results from VIS+NIR and literature values for $T_{\rm eff}$ (top panel), $\log{g}$ (middle panel), and [Fe/H] (bottom panel). The 1:1 relation is 
 indicated by the black line. The 
 uncertainties of this work (black) are shown in the lower right corner of each panel together with the uncertainties of \cite{Rajpurohit2018} (purple).}
 \label{fig:lit_VISNIR}
 \end{figure*}

We compare our results for different wavelength ranges to literature values \citet[][hereafter Mald15]{Maldonado2015}, RA12, GM14, \citet[][hereafter Mann15,]{Mann2015} and Raj2018. $T_{\rm eff}$ and 
metallicities of Mald15 were derived using pseudo-equivalent widths in optical spectra. They used photometric relations to 
derive stellar masses and obtained stellar radii from empirical mass-radius relations from interferometry \citep{vonBraun2014,Boyajian2012} and eclipsing binaries \citep{Hartman2015} to derive radii. From that 
they determined $\log{g}$ for their sample. Mann15 derived $T_{\rm eff}$ by fitting BT-Settl models \citep{Allard2013} to optical spectra. Metallicities were determined from NIR spectra using 
empirical relations between equivalent widths of atomic features and metallicity \citep{Mann2013,Mann2014}. In the latter, they obtained stellar masses using the mass-luminosity relation of 
\cite{Delfosse2000} and stellar radii by calculating the angular diameter from $T_{\rm eff}$ and the bolometric flux, and then by using the trigonometric parallax for each star. 
We calculated $\log{g}$ for the samples of GM14 and Mann15 from the stellar masses and radii they provided. Figures~\ref{fig:lit_VIS} -- \ref{fig:lit_VISNIR} show the comparisons between our 
parameters in the VIS, NIR, and VIS+NIR, respectively, and literature values for the stars in common. For better readability in the figures we plotted the uncertainties of our work and of 
Raj18 separately in the lower right corner of each panel. 

Temperatures derived by GM14 from VIS spectra are slightly higher than temperatures determined by Mann15, which are generally cooler than our results. This can be seen very well for our NIR spectra 
in Fig.~\ref{fig:lit_NIR} (top panel). Both authors used BT-Settl models to obtain $T_{\rm eff}$. Values from Raj18 show a large spread of sometimes up to 300~K compared to our results. Generally, our 
$T_{\rm eff}$ values for all wavelength ranges follow the literature very well. In $\log{g}$ and metallicity the plots look somewhat different. For both parameters, results from Raj18 do 
not correlate with our values nor with the other literature, spreading across the whole parameter range. For the other literature the plots are significantly different from the corresponding ones presented in 
Pass18, which can be explained by the use of different evolutionary and synthetic models, and the incorporation of different stellar ages. Concerning metallicity (bottom panels in Figs.~\ref{fig:lit_VIS} and
\ref{fig:lit_VISNIR}) we find a 
split relation, with the majority of our values being more metal-rich than the literature. We will discuss outliers in more detail on the basis of Fig.~\ref{fig:outliers_age} (top panel), where we 
include ages and activity in the plots. 

Since the comparisons of VIS, NIR, and VIS+NIR follow about the same pattern in Figs.~\ref{fig:lit_VIS} and \ref{fig:lit_VISNIR}, we will focus on the values of VIS+NIR for further discussion. Due to the 
large spread of results from Raj18 over the whole parameter range, we will exclude them from the following analysis.



We calculated the mean absolute difference (MAD) between our results in different wavelength ranges and literature values. The results are summarized in Table~\ref{tab:mad}. 
The highest MAD in temperature over all wavelength ranges is found for RA12. The other MADs correspond to our estimated uncertainties, some being only 10--20\,K higher. For $\log{g}$ the MADs are at least twice as high as our 
uncertainties. This is different for metallicity, where MADs for VIS+NIR and VIS lie mostly within our errorbars for these wavelength ranges. In the NIR the MADs are generally higher, which can be seen as well 
when comparing Figs.~\ref{fig:lit_VIS} and~\ref{fig:lit_NIR}, where the deviation from the 1:1 relation is larger in the NIR. Comparing MADs from Raj18 to those of other works shows much higher values, 
especially for $\log{g}$ and metallicity. Therefore, and for reference, we also calculated a total MAD excluding Raj18. 
In summary, it is shown that differences in the NIR for $\log{g}$ and metallicity are slightly higher than for VIS+NIR and VIS. 

 \begin{center}
   \begin{table*}
     \caption{Mean absolute difference between literature and results of this work for different wavelength ranges.}
     \label{tab:mad}
     \centering {
     \begin{tabular}{lccc}
       \hline \hline 
             
        Work& VIS+NIR & NIR & VIS \\
        & $T_{\rm eff}$ [K]/ $\log{g}$ / [Fe/H] & $T_{\rm eff}$ [K]/ $\log{g}$ / [Fe/H] & $T_{\rm eff}$ [K]/ $\log{g}$ / [Fe/H] \\
        \hline
        
        \cite{Maldonado2015} & 49.97 / 0.094 / 0.212 & 53.98 / 0.104 / 0.262 & 46.50 / 0.088 / 0.184 \\
        \cite{RojasAyala2012} & 84.02 / ... / 0.152 & 82.46 / ... / 0.194 & 82.03 / ... / 0.142 \\
        \cite{GaidosMann2014} & 68.85 / 0.095 / 0.157 & 57.64 / 0.109 / 0.228 & 72.51 / 0.097 / 0.133 \\
        \cite{Mann2015} & 60.19 / 0.119 / 0.152 & 66.07 / 0.137 / 0.226 &  55.88 / 0.119 / 0.136 \\
        \cite{Rajpurohit2018} & 77.19 / 0.401 / 0.248 & 86.63 / 0.423 / 0.289 & 69.87 / 0.397 / 0.246 \\
        \cite{Schweitzer2019} & 43.98 / 0.136 / 0.116 & 55.45 / 0.160 / 0.223 & 45.62 / 0.127 / 0.082 \\
        Total  & 61.63 / 0.223 / 0.173 & 68.79 / 0.244 / 0.245 & 59.27 / 0.219 / 0.155 \\ 
        Total \citep[w/o][]{Rajpurohit2018} & 53.69 / 0.125 / 0.134 & 59.62 / 0.144 / 0.222 & 53.86 / 0.119 / 0.108 \\
  
       \hline
     \end{tabular}
     }
   \end{table*}
 \end{center}

\subsection{Discussion of outliers}
\label{outliers}
\begin{figure*}[htb]
\centering
 \includegraphics[height=0.95\textheight]{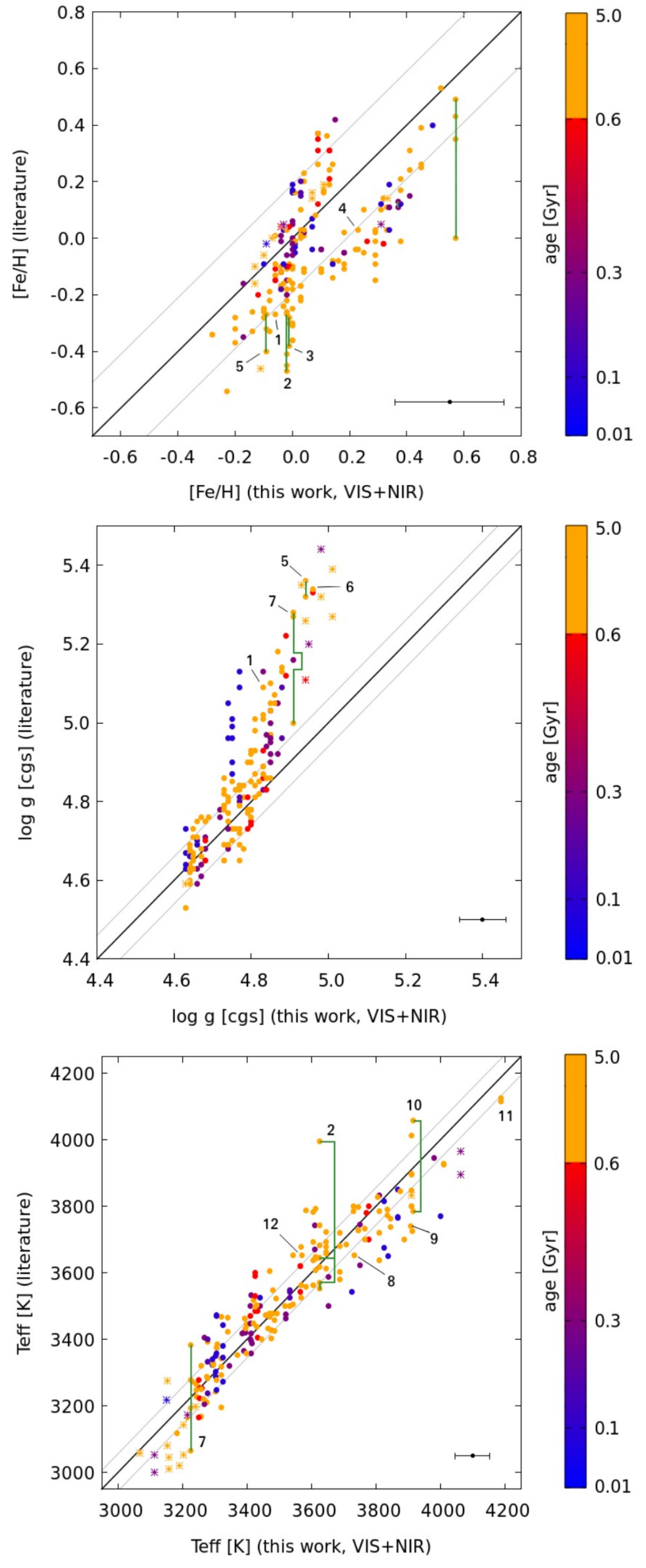}
 \caption{Comparison of [Fe/H] (top panel), $\log{g}$ (middle panel), and $T_{\rm eff}$ (bottom panel) between values of this work in VIS+NIR and literature. The age is color-coded;
 active star are  plotted as asterisks. Outliers are identified with numbers; the green lines connect their different literature values. The black line indicates the 1:1 relation, the gray lines the 
 1$\sigma$ deviation. }
 \label{fig:outliers_age}
 \end{figure*}

In the following we will discuss some outliers from Fig.~\ref{fig:lit_VISNIR}, mainly considering metallicity and $\log{g}$. Figure~\ref{fig:outliers_age} shows the same 
plots as Fig.~\ref{fig:lit_VISNIR}, with the estimated age of the stars color-coded. Additionally, 
active stars are plotted as asterisks. We define a star to be active if the H$\alpha$ pseudo-equivalent width is less than --0.3 \AA{} as described in \cite{Schoefer2019}, or if the star shows Ca~{\sc ii} 
emission (see Pass18). Active stars are identified in Table~\ref{tab:sample} with an activity flag 1. Furthermore, we define outliers if our value deviates by more than our estimated uncertainties 
from the literature values. For guidance, these 1$\sigma$ deviations are plotted as gray lines. 

First, we point out general trends seen in the plots. As mentioned above, the metallicity comparison shows a split correlation. In the top panel of Fig.~\ref{fig:outliers_age} we see that mainly stars 
located below the 1:1 relation have values deviating more than 1$\sigma$. Most active stars correspond very well to literature values, which supports our method of line selection since we found this 
parameter to be most influenced by activity. 

In $\log{g}$ (Fig.~\ref{fig:outliers_age}, middle panel) we find several young stars with lower values compared to literature. As explained above, they might still be contracting due to their young age of 
less than 50~Myr. 
Even though we carefully selected magnetically less sensitive lines, there are several stars in our sample that are considerably active, and therefore also less sensitive lines are affected. For metallicity 
(Fig.~\ref{fig:outliers_age}, top panel) all but a few of these star values coincide with the literature. However, in $\log{g}$ we can see them clearly as outliers at the upper end of the plot. They 
represent values mainly determined from masses and radii derived by Mann15. The same offset was also found by Schw19 for $\log{g}$ > 5.0. 
Additionally, we see many stars located above the 1:1 relation, meaning smaller $\log{g}$ compared to those from the literature. This might be explained by the use of the PARSEC evolutionary models, which provide 
smaller values than the Baraffe models used in Pass18 and Schw19. 
In the bottom panel of Fig.~\ref{fig:outliers_age} a small group of outliers at the cool end is represented by results from RA12 and Mann15, and was also found by Pass18. In addition we find these stars to 
be active. 

In the following discussion we will not include active stars and stars younger than 5~Gyr. For some stars our values derived from VIS, NIR, and VIS+NIR coincide within their errors, however they 
deviate by more than 1$\sigma$ from literature values. Their fits and spectra are of good or very good quality, therefore we are very confident of our resulting parameters and cannot find any explanation 
for their deviation from the literature. For this reason, these stars are not identified with a number and not considered as outliers. For stars with multiple literature values, we consider only those for 
which all literature values deviate more than 1$\sigma$ from our value to be 
outliers. For identification we connect these points using a green line. Also all stars listed in the following have very good 
quality spectra and fits in all wavelength ranges, unless stated otherwise. 
   \newline

\begin{itemize}
 
 \item[$\bullet$] (1) J12248-182, (2) J23492+024, (3) J11033+359: \newline
These stars appear more metal-poor in the NIR than in the VIS and VIS+NIR ranges. The NIR values coincide with literature values within 1$\sigma$. $T_{\rm eff}$ obtained from all three wavelength ranges 
agree with each other. Schw19 speculated if (1) was a subdwarf and therefore labeled it an outlier. (3) was also claimed to be an outlier in Schw19 due to a mismatch of photometric and interferometric 
radii. All three stars might be members of the thick disk according to their kinematics and therefore more than 5\,Gyrs old. 
\newline
 
 
 \item[$\bullet$] (4) J16581+257: \newline
 The VIS metallicity of this star is lower than those derived from NIR and VIS+NIR. However, with a value of +0.17\,dex it is still higher than literature values from RA12 (--0.04\,dex) 
 and Mann15 (+0.03\,dex). At this point we cannot explain the differences between our values and the literature.
  \newline 
  

 \item[$\bullet$] (5) J17578+046 (Barnard's star):\newline
 Since Barnard's star is claimed to be old \citep[7-10~Gyr][]{Ribas2018}, we assumed an age of 8~Gyr, using 
older evolutionary models resulting in smaller $\log{g}$. However, the final stellar parameters do not change significantly compared to the 5~Gyr model. $T_{\rm eff}$ and metallicities in all 
wavelength ranges coincide within their errors, but the metallicity seems to be a bit too high compared to the literature. Because the measurement of RA12 lies within 1$\sigma$, the star is not considered an 
outlier in metallicity. Assuming a more metal-poor value would translate into a \textasciitilde0.1~dex higher $\log{g}$, which would still be considerably smaller than literature values. 
\newline

 \item[$\bullet$] (6) J11509+483: \newline
 Although the quality of the spectrum and the fit for this star is moderate, the temperatures and metallicities obtained from different wavelength ranges coincide and are 
comparable with the literature (Mann15). One explanation for the deviation in $\log{g}$ (see Fig.~\ref{fig:outliers_age}, middle panel) might be that Mann15 determined too low of a stellar mass from the 
mass-luminosity relation \citep{Delfosse2000}, which results in too high of a $\log{g}$. 
Using an updated mass-luminosity relation \citep{Mann2019} Schw19 derived a slightly higher mass for this star. This would lead to a lower, but still too high $\log{g}$, which makes it more likely 
that the reason lies in the use of PARSEC models, where all $\log{g}$ values tend to be slightly smaller. \newline

\item[$\bullet$] (7) J11477+008: \newline
For this star there are literature values from Mald15, RA12, GM14, and Mann15. The star appears as an outlier only in $\log{g}$, where we measure values about 0.1--0.3\,dex smaller than in the literature. Our 
metallicity values in all wavelength ranges agree with the literature within their errors. RA12 measured a temperature about 200\,K cooler (3065\,K), whereas Mald15 determined a value of about 100\,K 
hotter (3382\,K). The other measurements agree very well with ours in all wavelength ranges. 
\newline

 \item[$\bullet$] (8) J22115+184: \newline
 Schw19 derived an almost 200~K cooler temperature for this star, as well as a considerably lower metallicity (--0.13\,dex). GM14 measured a temperature and 
metallicity of 3653~K and +0.26~dex, which both agree with our values from the VIS and VIS+NIR. In the NIR we measure a slightly higher temperature and considerably higher metallicity of +0.58\,dex. 
Since the spectrum and the fit are of very good quality, both in the VIS and NIR, we cannot pin down the reason for this discrepancy. \newline

 \item[$\bullet$] (9) J22503-070, (11) J04290+219, (12) J05127+196: \newline
 For these stars temperatures measured from the NIR correspond better to the literature, however they all lie within 100~K. \newline

 \item[$\bullet$] (10) J02222+478: \newline
 Literature values for this star are controversial, with 4058~K derived by GM14 and 3785~K obtained by Mann15. In the NIR we measured a temperature of 3840~K, which is about 
 100~K cooler than in VIS+NIR. However, our derived values lie well between the literature values, which is why we consider them as good measurements. 

\end{itemize}

\section{Summary}
The CARMENES instrument at Calar Alto Observatory performs high-accuracy radial velocity measurements in the VIS and NIR wavelength range simultaneously. We used the high-S/N template spectrum 
for each CARMENES sample M-dwarf to derive photospheric stellar parameters in the VIS, NIR, and VIS+NIR wavelength ranges for 282 stars. We calculated a new grid of PHOENIX model atmospheres incorporating a 
new equation of state, and new atomic and molecular line lists. Additionally, we carefully selected lines in the NIR 
that are sensitive to changes in stellar parameters, but insensitive to Zeeman-broadening caused by magnetic activity. Stellar activity is a crucial stellar property, influencing line profiles, which other 
studies did not consider. Furthermore, we involved different evolutionary models for deriving $\log{g}$ to account for stellar ages younger than 5~Gyr. 

For the first time we directly compared stellar parameters such as $T_{\rm eff}$, $\log{g}$, and [Fe/H] determined from multiple wavelength ranges simultaneously. 
There we find that $T_{\rm eff}$ is mainly consistent over all wavelength ranges, although we find a little larger offset in the NIR. This might be because the TiO-bands in the VIS are a very strong 
indicator for temperature. A direct comparison of $\log{g}$ shows that this parameter also corresponds very well in all wavelength ranges. A trend toward lower $\log{g}$ in the NIR can be found 
compared to values derived for the same stars in the VIS and VIS+NIR. Because the determination of $\log{g}$ depends on $T_{\rm eff}$ and metallicity, an explanation for that might be found 
in the 
metallicity. For this property we see a clear trend toward  more metal-rich values in the NIR than in the VIS. Also some values in VIS+NIR are slightly higher, which indicates a rise in metallicity 
toward longer wavelengths. Since metallicity is a basic property of the star and therefore wavelength independent, the cause for this most probably lies in the determination method or the synthetic models. 
Fitting synthetic spectra to derive stellar parameters has already been used in several studies \citep[e.g.,][]{GaidosMann2014,LindgrenHeiter2017,Passegger2016,Passegger2018,Rajpurohit2018}. While the fitting 
process itself is less crucial for the final result, the use of different synthetic model grids could lead to significant differences. However, a discussion on this topic is beyond the scope of this work. 

As mentioned by other works before \citep[e.g.,][]{Passegger2016, Passegger2018} synthetic model spectra considerably improved over the last years, however they still suffer from some deficiencies. In the 
VIS, these shortcomings have been pointed out by \cite{Passegger2018}, concerning Ti~{\sc i} ($\lambda$ 846.9 and 867.77~nm) and Fe~{\sc i} ($\lambda$ 867.71 and 882.6~nm). In the NIR we find similar 
deficiencies in the K~{\sc i} lines where the models cannot fit the line cores. Moreover, due to the different spatial radial velocity and heliocentric correction for each star, the Fe~{\sc i} line at 
$\lambda$ 1112.3~nm is for some stars contaminated by telluric lines and therefore cannot be used. 

Comparisons of our derived parameters to literature values shows good agreement for the effective temperature. Although most of our values are consistent with the literature in $\log{g}$ and metallicity, 
there is a trend toward lower $\log{g}$ (most likely due to the use of PARSEC evolutionary models) and higher metallicity. As for now we cannot be absolutely certain about the reason for finding more 
metal-rich values in the NIR. 
Results from \cite{Rajpurohit2018} exhibit a large spread over the whole parameter range, with high deviations of up to 400~K in temperature and 1.0~dex in $\log{g}$ and metallicity compared to our work. We 
cannot find any correlations with our values or other literature. The overall distribution of the \cite{Rajpurohit2018} parameters is similar to what we found in early studies on stellar parameter 
determination by leaving $\log{g}$ as a free fit parameter. That method might reduce biases in the parameter space, but it also is more susceptible to degeneracies between the parameters, especially 
$\log{g}$ and metallicity, as described in \cite{Passegger2018}.

Precise determination of the metallicity is still a challenging task. Different methods and synthetic models can lead to different results. In the literature we sometimes find very diverse measurements for 
the same star (e.g., GJ~205: 0.0 \cite{Maldonado2015}, +0.49 \cite{Mann2015}). 
Our comparison of stellar parameters determined from multiple wavelength ranges shows that deviations from the literature are smallest for the VIS, followed by the VIS+NIR, and highest in the NIR, 
especially concerning metallicity. This might be explained by continuing shortcomings in synthetic models and the lower number of useful and parameter-sensitive lines in the NIR compared to the VIS. 
However, the differences between VIS and VIS+NIR are marginal. For that reason, we emphasize the use of both ranges for parameter determination in order to maximize the amount of spectral information 
available and minimize possible effects caused by imperfect modeling.



\begin{acknowledgements}
We thank an anonymous referee for helpful comments that improved the quality of this paper.
CARMENES is an instrument for the Centro Astron\'omico Hispano-Alem\'n de Calar Alto (CAHA, Almer\'ia, Spain). CARMENES
is funded by the German Max-Planck-Gesellschaft (MPG), the Spanish Consejo Superior de Investigaciones Cient\'ificas (CSIC),
the European Union through FEDER/ERF FICTS-2011-02 funds, and the members of the CARMENES Consortium (Max-
Planck-Institut f\"ur Astronomie, Instituto de Astrof\'isica de Andaluc\'ia, Landessternwarte K\"onigstuhl, Institut de Ci\`ncies de
l'Espai, Insitut f\"ur Astrophysik G\"ottingen, Universidad Complutense de Madrid, Th\"uringer Landessternwarte Tautenburg,
Instituto de Astrof\'isica de Canarias, Hamburger Sternwarte, Centro de Astrobiolog\'ia, and Centro Astron\'omico Hispano-
Alem\'an), with additional contributions by the Spanish Ministry of Economy, the German Science Foundation through the
Major Research Instrumentation Programme, and DFG Research Unit FOR2544 “Blue Planets around Red Stars”, the Klaus
Tschira Stiftung, the states of Baden-W\"urttemberg and Niedersachsen, and by the Junta de Andaluc\'ia.
This work is based on data from the CARMENES data archive at CAB (INTA-CSIC).
We acknowledge financial support from the Agencia Estatal de Investigaci\'on
of the Ministerio de Ciencia, Innovaci\'on y Universidades, and the European
FEDER/ERF funds through projects AYA2018-84089, ESP2017-87676-C5-1-R, ESP2016-80435-C2-1-R, AYA2016-79425-C3-1/2/3-P, AYA2015-69350-C3-2-P, and the Centre of Excellence ''Severo
Ochoa'' and ''Mar\'ia de Maeztu'' awards to the Instituto de Astrof\'isica de
Canarias (SEV-2015-0548), Instituto de Astrof\'isica de Andaluc\'ia
(SEV-2017-0709), and Centro de Astrobiolog\'ia (MDM-2017-0737), and the
Generalitat de Catalunya/CERCA programme''.
Some of the  calculations presented here were performed at the RRZ of the Universit\"at Hamburg, at the H\"ochstleistungs 
Rechenzentrum Nord (HLRN), and at the National Energy Research Supercomputer Center (NERSC), which is supported by the 
Office of Science of the U.S. Department of Energy under Contract No. DE-AC03-76SF00098.  We thank all these 
institutions for a generous allocation of computer time. PHH gratefully acknowledges the support of NVIDIA
Corporation with the donation of a Quadro P6000 GPU used in this research. 
This work has made use of the VALD database, operated at Uppsala University, the Institute of Astronomy RAS in Moscow, and the University of Vienna.
This work has made use of data from the European Space Agency (ESA) mission
{\it Gaia} (\url{https://www.cosmos.esa.int/gaia}), processed by the {\it Gaia}
Data Processing and Analysis Consortium (DPAC,
\url{https://www.cosmos.esa.int/web/gaia/dpac/consortium}). Funding for the DPAC
has been provided by national institutions, in particular the institutions
participating in the {\it Gaia} Multilateral Agreement.

\end{acknowledgements}
\clearpage
\bibliographystyle{aa}  
\bibliography{CARM_NIR.bib}

\newpage
\appendix
\section{Additional plots}

\begin{figure}[htb]
\centering
 \includegraphics[height=0.85\textheight]{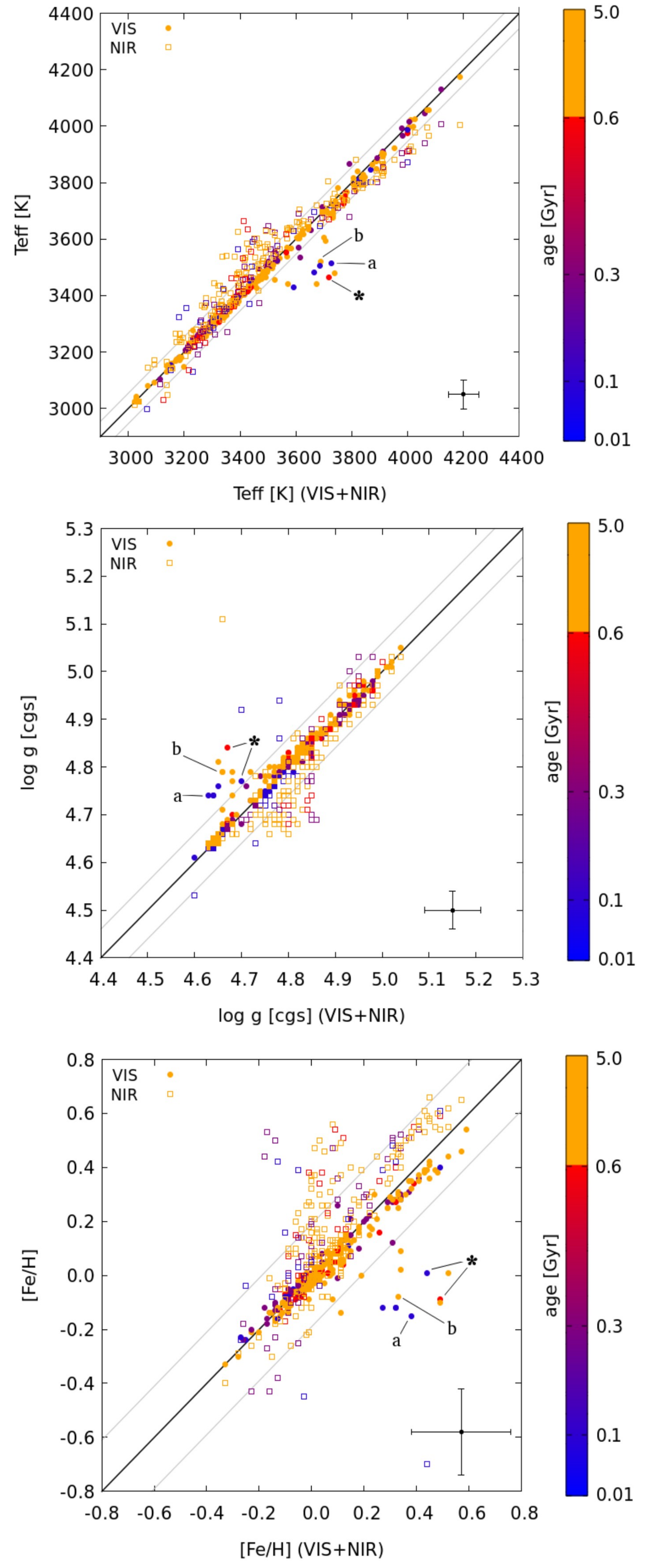}
 \caption{Comparison of stellar parameters derived from different wavelength ranges. Values from VIS+NIR are plotted on the x-axis, values for VIS (filled circles) and 
 NIR (open squares) are shown on the y-axis with the age color-coded. The black line indicates the 1:1 relation, the gray lines represent the 1$\sigma$ deviations for VIS+NIR of 54~K, 0.06~$\log{g}$, and 
 0.19~[Fe/H].}
 \label{fig:comparison_VIS+NIR}
 \end{figure}

Figure~\ref{fig:comparison_VIS+NIR} presents the same comparison plots as in Fig.~\ref{fig:comparison}, including values for the combined wavelength ranges VIS+NIR. The behavior of the NIR values compared to 
VIS+NIR is very similar to what was already shown in Fig.~\ref{fig:comparison}, where we compared the NIR with the VIS results. A small group of outliers from the VIS is found between 3600\,K and 3800\,K 
that exhibit about 200\,K cooler temperatures compared to the NIR and VIS+NIR. About half of them have ages younger than 0.15\,Gyr. The same stars are also outliers in $\log{g}$ (middle panel) and 
metallicity (bottom panel). 

For $\log{g}$ (middle panel) it is shown that most values coincide very well over all wavelength ranges. The small group of outliers in the VIS can be found with $\log{g}$ of about 0.1 -- 0.2\,dex higher 
compared to the NIR and VIS+NIR.

The comparison in metallicity is presented in the bottom panel of Fig.~\ref{fig:comparison_VIS+NIR}. The VIS outlier group exhibits up to 0.4\,dex lower metallicities compared to VIS+NIR, whereas values for 
other stars perfectly agree in VIS and VIS+NIR. 
We will discuss the properties of the VIS outlier group in more detail in the following.

\subsection*{The VIS outlier group}
This group consists of nine stars, for which we derive cooler $T_{\rm eff}$, larger $\log{g}$, and lower metallicities compared to the NIR and VIS+NIR. Hence, their parameters in the latter two 
wavelength ranges agree with each other within 1$\sigma$. Four of these stars are younger than 0.15\,Gyr and two are magnetically active (identified with an asterisk in Fig.~\ref{fig:comparison_VIS+NIR}). 
However, all fits to the spectra in either wavelength range are of good or very good quality, which cannot explain the deviations in the parameters. For two stars, which we marked with letters 
$a$ and $b$, there are literature values available.\newline 
\begin{itemize}
\item[(a)] J02358+202: Literature values for $T_{\rm eff}$ and $\log{g}$ from Mann15 perfectly agree with our VIS values, however, they measured a metallicity of +0.12\,dex, 
which lies between our values for VIS (--0.15\,dex) and VIS+NIR (+0.38\,dex). The estimated age for this star is 0.1\,Gyr.
\item[(b)] J04429+189: For these star there are four different literature references from Mald15, RA12, GM14, and Mann15. Temperatures are spread between 
3581\,K (RA12) and 3721\,K (GM14). Metallicities range from +0.03\,dex (Mald15) and +0.14\,dex (GM14, Mann15) and lie between our measurements in the VIS (--0.08\,dex) and VIS+NIR (+0.33\,dex). 
\end{itemize}

\clearpage
\section{Online material}
These tables are available in their entirety in a machine-readable form at the CDS via anonymous ftp to cdsarc.u-strasbg.fr (130.79.128.5) or via 
http://cdsweb.u-strasbg.fr/cgi-bin/qcat?J/A+A/. An excerpt is shown here for guidance regarding their form and content.

\begin{longtable}{lllcccccc}
\label{tab:sample}\\
\caption[]{CARMENES sample stars investigated in this work$^a$. }\\
   \hline
   \hline
   \noalign{\smallskip}
Karmn  & Name & GJ & $\alpha$ [deg] & $\delta$ [deg] 	&  Spectral	 & $v\sin{i}$& Assumed  & Activity\\ 
     & &  & & & type	& [km\,s$^{-1}$]& age [Gyr] & flag\\
\noalign{\smallskip}
    \hline
    \noalign{\smallskip}		
 \endfirsthead
\caption[]{CARMENES sample stars investigated in this work$^a$.}\\ 
  \hline
  \hline
  \noalign{\smallskip}		
Karmn  & Name & GJ & $\alpha$ [deg] & $\delta$ [deg] &  Spectral	& $v\sin{i}$ & Assumed 	& Activity\\ 
   & &  & & & type	& [km\,s$^{-1}$] & age [Gyr] & flag\\
  \noalign{\smallskip}
  \hline
  \noalign{\smallskip}
  \endhead
  \noalign{\smallskip}
  \hline
  \endfoot
J00051+457	&	GJ	2	&	2	&	00:05:12.18	&	+45:47:09.3	&	M1.0\,V	&	 $\le$ 2.0 	&	0.60	& 1	\\
J00067--075	&	GJ 1002		&	1002	&	00:06:42.35	&	-07:32:46.4	&	M5.5\,V	&	 $\le$ 2.0 	&	5.00	& 0	\\
J00162+198E	&	LP 404-062		&	1006B	&	00:16:16.93	&	+19:51:38.9	&	M4.0\,V	&	 $\le$ 2.0 	&	5.00	& 0	\\
J00183+440	&	GX	And	&	15A	&	00:18:27.04	&	+44:01:29.0	&	M1.0\,V	&	 $\le$ 2.0 	&	5.00	& 0	\\
J00184+440	&	GQ	And	&	15B	&	00:18:29.94	&	+44:01:43.3	&	M3.5\,V	&	 $\le$ 2.0 	&	0.60	& 0	\\
J00286--066	&	GJ	1012	&	1012	&	00:28:39.12	&	-06:40:01.6	&	M4.0\,V	&	 $\le$ 2.0 	&	5.00	& 0	\\
J00389+306	&	Wolf	1056	&	26	&	00:39:00.91	&	+30:36:58.8	&	M2.5\,V	&	 $\le$ 2.0 	&	5.00	& 0	\\
J00570+450	&	G	172-030	&	...	&	00:57:03.61	&	+45:05:08.7	&	M3.0\,V	&	 $\le$ 2.0 	&	0.60	& 0	\\
\end{longtable}

\begin{flushleft}
$^a$ Carmencita identifier, recommended name, Gliese-Jahreiss number, {\em Gaia} DR2 equatorial coordinates, spectral type, rotational velocity from \cite{Reiners2018a} except for 
J10196+198 (AD Leo), which is from \cite{Martinez2014}, assumed age from kinematics, and an activity flag (1 if active).
\end{flushleft}



 \begin{landscape}
\begin{longtable}{lcccccccccc}
\label{tab:results}\\
\caption[]{Basic astrophysical parameters of investigated stars for different wavelength regimes.}\\
   \hline
   \hline
   \noalign{\smallskip}
  & \multicolumn{3}{c}{VIS+NIR} & \multicolumn{3}{c}{NIR} & \multicolumn{3}{c}{VIS} \\
Karmn  &  $T_{\rm eff}$ & $\log{g}$ & [Fe/H]	& $T_{\rm eff}$ & $\log{g}$ & [Fe/H]	& $T_{\rm eff}$ & $\log{g}$ & [Fe/H]	\\ 
\noalign{\smallskip}
    \hline
    \noalign{\smallskip}		
 \endfirsthead
\caption[]{Basic astrophysical parameters of investigated stars (cont.).}\\ 
  \hline
  \hline
  \noalign{\smallskip}		
  & \multicolumn{3}{c}{VIS+NIR} & \multicolumn{3}{c}{NIR} & \multicolumn{3}{c}{VIS} \\
Karmn  & $T_{\rm eff}$ & $\log{g}$ & [Fe/H]	& $T_{\rm eff}$ & $\log{g}$ & [Fe/H]	& $T_{\rm eff}$ & $\log{g}$ & [Fe/H]	\\ 
  \noalign{\smallskip}
  \hline
  \noalign{\smallskip}
  \endhead
  \noalign{\smallskip}
  \hline
  \endfoot
J00051+457	&	3772	$\pm$	54	&	4.68	$\pm$	0.06	&	+0.26	$\pm$	0.19	&	3738	$\pm$	56	&	4.68	$\pm$	0.04	&	+0.35	$\pm$	0.16	&	3726	$\pm$	51	&	4.70	$\pm$	0.04	&	+0.16	$\pm$	0.16	\\
J00067--075	&	3038	$\pm$	54	&	5.04	$\pm$	0.06	&	--0.10	$\pm$	0.19	&	3024	$\pm$	56	&	5.03	$\pm$	0.04	&	--0.03	$\pm$	0.16	&	3023	$\pm$	51	&	5.05	$\pm$	0.04	&	--0.08	$\pm$	0.16	\\
J00162+198E	&	3272	$\pm$	54	&	4.85	$\pm$	0.06	&	+0.11	$\pm$	0.19	&	3360	$\pm$	56	&	4.80	$\pm$	0.04	&	+0.11	$\pm$	0.16	&	3261	$\pm$	51	&	4.86	$\pm$	0.04	&	+0.08	$\pm$	0.16	\\
J00183+440	&	3606	$\pm$	54	&	4.77	$\pm$	0.06	&	--0.14	$\pm$	0.19	&	3628	$\pm$	56	&	4.77	$\pm$	0.04	&	--0.18	$\pm$	0.16	&	3601	$\pm$	51	&	4.78	$\pm$	0.04	&	--0.16	$\pm$	0.16	\\
J00184+440	&	3261	$\pm$	54	&	4.96	$\pm$	0.06	&	--0.12	$\pm$	0.19	&	3264	$\pm$	56	&	4.94	$\pm$	0.04	&	--0.05	$\pm$	0.16	&	3256	$\pm$	51	&	4.97	$\pm$	0.04	&	--0.13	$\pm$	0.16	\\
J00286--066	&	3343	$\pm$	54	&	4.80	$\pm$	0.06	&	+0.15	$\pm$	0.19	&	3417	$\pm$	56	&	4.75	$\pm$	0.04	&	+0.23	$\pm$	0.16	&	3321	$\pm$	51	&	4.83	$\pm$	0.04	&	+0.09	$\pm$	0.16	\\
J00389+306	&	3491	$\pm$	54	&	4.78	$\pm$	0.06	&	--0.02	$\pm$	0.19	&	3571	$\pm$	56	&	4.69	$\pm$	0.04	&	+0.26	$\pm$	0.16	&	3482	$\pm$	51	&	4.79	$\pm$	0.04	&	--0.05	$\pm$	0.16	\\
J00570+450	&	3397	$\pm$	54	&	4.85	$\pm$	0.06	&	--0.01	$\pm$	0.19	&	3414	$\pm$	56	&	4.86	$\pm$	0.04	&	--0.05	$\pm$	0.16	&	3400	$\pm$	51	&	4.85	$\pm$	0.04	&	--0.01	$\pm$	0.16	\\
\end{longtable}

\begin{flushleft}
\end{flushleft}

 \end{landscape}


\end{document}